\documentstyle[12pt]{article}
\begin{document}
\baselineskip6mm
\newcommand{\x}{{\underline x}}
\newcommand{\omfett}{\mbox{\boldmath $\omega$}}
\newcommand{\efett}{\mbox{\boldmath $e$}}
\newcommand{\gfett}{\mbox{\boldmath $g$}}
\newcommand{\alfett}{\mbox{\boldmath $\alpha$}}
\newcommand{\de}{\delta}
\newcommand{\ga}{\gamma}
\newcommand{\e}{\epsilon}
\newcommand{\th}{\theta}
\newcommand{\ot}{\otimes}
\newcommand{\ba}{\begin{array}} \newcommand{\ea}{\end{array}}
\newcommand{\beq}{\begin{equation}}\newcommand{\eeq}{\end{equation}}
\newcommand{\tmod}{{\cal T}}\newcommand{\amod}{{\cal A}}
\newcommand{\bemod}{{\cal B}}\newcommand{\cmod}{{\cal C}}
\newcommand{\dmod}{{\cal D}}\newcommand{\hmod}{{\cal H}}
\newcommand{\s}{\scriptstyle}
\newcommand{\tr}{{\rm tr}}
\newcommand{\einsop}{{\bf 1}}
\newcommand{\tnull}{\tmod_{\s 0}}
%
\title{Highest Weight $U_q[sl(n)]$ Modules\\
 and Invariant Integrable $n$-State Models\\
 with Periodic Boundary Conditions}
\author{A. Zapletal
\thanks{Supported by DFG, Sonderforschungsbereich 288 'Differentialgeometrie
        und Quantenphysik'}
\thanks{e-mail: zapletal@omega.physik.fu-berlin.de}
\\M. Karowski}
\date{\small\it Institut f\"{u}r Theoretische Physik\\
      Freie Universit\"{a}t Berlin\\Germany}
\maketitle
\begin{abstract}
The weights are computed for the Bethe vectors of an RSOS type
model with periodic boundary conditions obeying $U_q[sl(n)]$
($q=\exp(i\pi/r)$) invariance.
They are shown to be highest weight vectors.
The q-dimensions of the corresponding irreducible representations
are obtained.
\end{abstract}
%
%
In the last years considerable progress has been made on
the ''quantum symmetry" of integrable quantum chain models as the
XXZ-Heisenberg model and its generalizations. In \cite{p1} we constructed
an $sl_q(n)$ invariant RSOS type model with periodic boundary conditions.
In the present paper we prove for this model
the highest weight property of the Bethe states,
calculate the weights and the q-dimensions
of the representations and classify the irreducible ones.
For the case of open boundary conditions see e.g.~\cite{xxzopen},
\cite{devegahw} and \cite{georg}.

The model of \cite{p1} is defined by the transfer matrix
$\tau=\tau^{(n)}$ where
\beq
\label{transfer}
\tau^{(k)}(x,\x^{(k)})=\tr_q({\cal T}^{(k)}(x,\x^{(k)})=\sum_\alpha
q^{n+1-2\alpha}({\cal T}^{(k)})_\alpha^\alpha(x,\x^{(k)}),\quad
k=1,\dots,n.
\eeq
The ``doubled" monodromy matrix is given by
\beq
\label{monodromy}
{\cal T}^{(k)}_0(x,\x^{(k)})=\tilde T^{(k)}_0\cdot T^{(k)}_0(x,\x^{(k)})=
(R_{01}\dots R_{0N_k})\cdot(R_{N_k0}(x_{N_k}/x^{(k)})\dots
R_{10}(x_1/x^{(k)})).
\eeq
For the purpose of the nested algebraic Bethe ansatz in addition to
${\cal T}(x)={\cal T}^{(n)}(x)$ the monodromy matrices
for all $k\le n$ are needed.
The $sl_q(k)$ R-matrix is given by
\beq
\label{rmatrix}
R(x)=xR-x^{-1}PR^{-1}P,\quad
R=
\sum_{\alpha\neq \beta}E_{\alpha\alpha}\ot E_{\beta\beta}+q\sum_{\alpha}
E_{\alpha\alpha}\ot E_{\alpha\alpha}+(q-q^{-1})
\sum_{\alpha>\beta}E_{\alpha\beta} \ot E_{\beta\alpha},
\eeq
The Yang-Baxter equation reads
\beq
\label{ybe}
R_{12}(y/x){\cal T}_1(x)R_{21}{\cal T}_2(y)=
{\cal T}_2(y)R_{12}{\cal T}_1(x)R_{21}(y/x).
\eeq
The model is quantum group invariant, i.e. the transfer matrix
commutes with the generators of $U_q[sl(n)]$. These are obtained from
the monodromy matrices $T(x)$ in the limits
$x$ to $0$ or $\infty$ (up to normalizations)
\beq
\label{generator1}
T=\left( \ba{cccc}
	1&0&\cdots &0\\
	\alpha E_1 &1&\ddots&\vdots\\
	&\ddots &\ddots & 0 \\
	{*} & &\alpha E_{n-1}&1
\ea\right) q^{\bf W},\quad
T_\infty=q^{-\bf W}
\left(\ba{cccc} 1&-\alpha F_1 & &*\\
0&1&\ddots& \\&\ddots&\ddots&-\alpha F_{n-1}\\
0&&0&1\ea\right),
\eeq
where $\alpha=q-q^{-1}$ and the matrix
${\bf W}=\mbox{diag}\{W_1,\ldots,W_n\}$ contains the $U_q[gl(n)]$
Cartan elements.
Analogously to eq.~(\ref{generator1}), we introduce
as a limit of ${\cal T}^{(n)}(x,\x^{(n)})$ for $x$ to $0$
\beq
\label{generator2}
{\cal T}=\tilde T\cdot T,\quad{\rm where}\quad \tilde T=T^{-1}_\infty,
\eeq
here and in the following operators without argument denote these limits
for $x$ to $0$.\\
We write the doubled monodromy matrices as
$k\times k$ block-matrices of operators
\beq
\label{block}
{\cal T}^{(k)}(x)=\left( \ba{ll} \amod^{(k)}(x) & \bemod^{(k)}(x) \\
\cmod^{(k)}(x) & \dmod^{(k)}(x) \ea \right).
\eeq
We also introduce the
reference states $\Phi^{(k)}$ with $\cmod^{(k)}(x)\Phi^{(k)}=0$.
The eigenstates of the transfer matrix $\tau(x)$ are the Bethe ansatz states
$\Psi=\Psi^{(n)}$ obtained by the nested procedure
\beq
\label{bethevector}
\Psi^{(k)}=\bemod^{(k)}_{\alpha_1}(x^{(k-1)}_1)\dots
\bemod^{(k)}_{\alpha_{N_{k-1}}}(x^{(k-1)}_{N_{k-1}})\Phi^{(k)}
\Psi^{(k-1)}_{\underline \alpha},\quad (k=2,\dots,n),\
\Psi^{(1)}=1.
\eeq
The sets of parameters $x_j^{(k)}=\exp\left(i\theta_j^{(k)}-(n-k)
\gamma/2\right)$ $(q=e^{i\gamma})$ fulfil the Bethe ansatz equations:
for $j=1,\dots,N_k$ and $k=1,\dots,n-1$
\beq
\label{bae}
 q^{2+w_{n-k}-w_{n-k+1}}\prod_{l=k \pm 1}\prod_{i=1}^{N_{l}}
\frac{\sinh\frac{1}{2}\left(\theta_j^{(k)}-\theta_i^{(l)}-i\gamma\right)}
{\sinh\frac{1}{2}\left(\theta_j^{(k)}-\theta_i^{(l)}+i\gamma\right)}
\prod_{i=1}^{N_{k}}
\frac{\sinh\frac{1}{2}\left(\theta_j^{(k)}-\theta_i^{(k)}+2i\gamma\right)}
{\sinh\frac{1}{2}\left(\theta_j^{(k)}-\theta_i^{(k)}-2i\gamma\right)}=-1
\eeq
where, below, the $w_i=N_{n-i+1}-N_{n-i}$ will turn out to be the
weights of the state $\Psi$, i.e.~the eigenvalues of the $W_i$'s defined
by eq.~(\ref{generator1}).

\bigskip
\noindent{\bf Theorem}: {\em
The Bethe ansatz states are highest weight states, i.e.}
\[ E_i\psi=0\quad (i=1,\dots,n-1). \]
The analogous statement for the case of open boundary conditions has been
proved in [3].

\medskip
\noindent\underbar{\it Proof}:
First we prove ${\cal T}^\alpha_\beta\Psi=0$
and then  $T^\alpha_\beta\Psi=0$ for $\alpha >\beta$.
The Yang-Baxter relation (\ref{ybe}) implies
\beq
\label{com1}
\tmod^\alpha_\beta \bemod_\gamma(x)=
\bemod_{\gamma'}(x)\tmod^{\alpha'}_{\beta'}
R^{\gamma' \alpha}_{\gamma'' \alpha'} R^{\beta' \gamma''}_{\beta\gamma} ,
\quad \mbox{(for $\alpha>1$, else see eq.~(\ref{cvert}))}.
\eeq
We apply the technique of the nested algebraic Bethe ansatz and commute
${\cal T}$ through all the ${\cal B}$'s of  $\Psi$ in eq.~(\ref{bethevector})
\beq
\label{com2}
\tmod^\alpha_\beta
\bemod_{\alpha_1}(x_1)\dots\bemod_{\alpha_{N_{n-1}}}(x_{N_{n-1}})\Phi^{(k)}
\Psi^{(n-1)}_{\underline \alpha}=
\bemod_{\alpha_1}(x_1)\dots\bemod_{\alpha_{N_{n-1}}}(x_{N_{n-1}})\Phi^{(k)}
\left(\tmod^{(n-1)}\right)^\alpha_\beta\Psi^{(n-1)}_{\underline \alpha}.
\eeq
Iterating this procedure $\beta-1$ times we arrive at
${\cal C}^{(k)}_{\alpha'}\Psi^{(k)}$ with $k=n-\beta+1$ and
$\alpha'=\alpha-\beta$.
At this stage we use, as usual (see e.g.~ref.~\cite{devegan}),
the commutation rule
\beq
\label{cvert}
\cmod^{(k)}_{\alpha'}\bemod^{(k)}_{\gamma}(x)=q^{-1}R^{\ga' \alpha'}
_{\ga \alpha''} \bemod^{(k)}_{\ga'}(x) \cmod^{(k)}_{\alpha''}
+(1-q^{-2})\left((\dmod^{(k)})^{\alpha'}_{\gamma}\amod^{(k)}(x)-
(\dmod^{(k)})^{\alpha'}_{\alpha''}(\dmod^{(k)})^{\alpha''}_{\gamma}(x)\right),
\eeq
to prove that the Bethe ansatz equations (\ref{bae}) imply
\beq
\label{zero1}
{\cal C}^{(k)}_{\alpha'}\Psi^{(k)}=0\quad (k=2,\dots,n)
\quad{\rm and\ therefore}\quad
{\cal T}^\alpha_\beta\Psi=0 \quad{\rm for}\quad \alpha >\beta.
\eeq
Finally we show $T^\alpha_\beta\Psi=0$ for $\alpha >\beta$.
We have with eqs.~(\ref{generator1}) and (\ref{zero1}) for all $\beta<n$
\beq
\label{zero2}
\tmod^n_\beta\Psi = \tilde{T}^n_nT^n_\beta\Psi =0
\eeq
and because
$\tilde{T}^n_n$ is an invertible operator it follows that
$T^n_\beta\Psi=0$. Now we consider the previous row, where $\beta<n-1$:
\beq
\label{zero3}
\tmod^{n-1}_\beta\Psi=\left(\tilde{T}^{n-1}_{n-1}T^{n-1}_{\beta}+
\tilde{T}^{n-1}_{n}(T_0)^{n}_{\beta} \right)\Psi=0,
\eeq
and therefrom along with the foregoing result we get $(T_0)^{n-1}_\beta
\Psi=0$.
By iteration we find $T^\alpha_\beta\Psi=0$ for $\alpha >\beta$ and
hence from eq.~(\ref{generator1}) $E_i\Psi=0$ for all
$i$, this proves the highest weight property of the Bethe vectors.\par

Next we compute the weights of the Bethe ansatz states $\Psi$.
We consider first
%
\beq \label{decomp}
\tmod^\alpha_\alpha\Psi=\tilde{T}^\alpha_\alpha T^\alpha_\alpha\Psi+
\sum_{\beta>\alpha}
\tilde{T}^\alpha_\beta T^\beta_\alpha \Psi.
\eeq
Again shifting $\tmod^\alpha_\alpha$ to the right as in eq.~(\ref{com2})
we get the operator $(\tmod^{(n-1)})^\alpha_\alpha$.
By iteration we arrive at $\amod^{(k)} \Psi^{(k)}$ for $k=n-\alpha+1$.
The Yang-Baxter relation (\ref{ybe}) and eqs.~(\ref{monodromy}) and
(\ref{rmatrix}) imply
\beq
\label{weight1}
{\cal A}^{(k)}{\cal B}^{(k)}_\beta(x)=
q^{-2}{\cal B}^{(k)}_\beta(x){\cal A}^{(k)},\quad
{\cal A}^{(k)}\Psi^{(k)}=q^{2N_{k}}\Psi^{(k)}
\eeq
and therefore
\beq
\label{weight2}
\amod^{(k)} \Psi^{(k)}=q^{2(N_{k}-N_{k-1})}\Psi^{(k)}.
\eeq
 From eqs.~(\ref {generator1}) and (\ref{generator2}) we have
\beq
\label{weight3}
\tilde{T}^i_i=T^i_i=q^{W_i}\quad{\rm and}\quad{\cal T}^i_i=q^{2W_i}.
\eeq
and finally
\beq
\label{weight4}
q^{2W_i}\Psi=q^{2w_i}\Psi\quad{\rm with}\quad
w_i= N_{n-i+1}-N_{n-i}.
\eeq
So any Bethe ansatz solution is characterized by a weight vector
\beq \label{weightvector}
w=(w_1,\dots,w_n)=(N_{n}-N_{n-1},\dots,N_{2}-N_{1},N_{1})
\eeq
with the usual highest weight condition
\beq \label{hw}
w_1\ge\cdots\ge w_n\ge 0.
\eeq
Here $N=N^{(n)}$ is the number of lattice sites and $N^{(k)}$
($k=n-1,\dots,1$) is the
number of roots in the $k$-th Bethe ansatz level.
The highest weight condition (\ref{hw}) may be shown as usual.
The result (\ref{weightvector}) is consistent with the ``ice rule"
fulfilled by the $R$-matrix
(\ref{rmatrix}). This means that each operator ${\cal B}^{(k)}_\alpha(x)$
reduces $w_k$ and lifts $w_\alpha$ by one.

The q-dimension of a representation $\pi$ with representation space $V$
is obtained from the ``Markov trace" (see e.g.~ref.~\cite{resht})
\beq \label{qdim}
\dim_q \pi=\tr_V \left(q^{-\sum_{i>j}(W_j-W_i)} \right).
\eeq
As is well known for the case of $q$ being a root of unity
the generators $E_i$ and $F_i$ become nilpotent
\beq\label{nil}
(E_i)^r=(F_i)^r=0,\quad q=\exp(i\pi/r),\quad r=n+2,n+3,\ldots.
\eeq
A highest weight module is equivalent to the corresponding one of $sl(n)$,
if this relation does not concern it.
These representations remain still irreducible and will be called good ones.
The other representations are called bad and up to special irreducible
cases with vanishing q-dimension
they are reducible but not decomposable.

For the irreducible representations $\pi_w$ with
highest weight
vector $w$ eq.~(\ref{qdim}) gives
\beq
\label{qdim1}
\dim_q \pi_w=   
\prod _{\alpha \in \Phi_+} \frac
{\left[(w+g,\alpha)\right]_q}{\left[(g,\alpha)\right]_q}=
\prod_{i>j}\frac{\left[w_j-w_i+i-j\right]_q}{\left[
i-j \right]_q},
\eeq
where $\quad [x]_q=(q^x-q^{-x})/(q-q^{-1})$ is a q-number,
$\Phi_+$ denotes the set of positive roots and $g$ is the
Weyl vector $g=\frac{1}{2} \sum_{\alpha \in \Phi_+} \alpha$.
The good representations are characterized by positive q-dimensions.
{}From eq.~(\ref{qdim1}) it follows that their weight patterns are restricted
in their length
\beq
w_1-w_n \leq r-n,
\eeq
The q-dimensions of bad representations vanish.

It is an interesting question, how these good representations are
characterized in the language of the Bethe ansatz.
In ref.~\cite{georg}
it is shown for $sl_q(2)$ that these are given by all Bethe ansatz solutions
with only positive parity strings (in the language of Takahashi
\cite{takahashi}).
In a forthcoming paper we will show
how this classification extends to q-symmetries of higher rank.
%
%
%
%
%

%

\begin{thebibliography}{99}
%
\bibitem{p1} M. Karowski and A. Zapletal, Nucl. Phys.
{\bf B} (1994) in press.
%
\bibitem{xxzopen}
L. Mezincescu and R.I. Nepomechie, Mod. Phys. Lett. {\bf A6} (1991)
2497;\\
C. Destri and H.J. de Vega, Nucl. Phys. {\bf B361}
(1992) 361;\\
A. F\"orster and M. Karowski, Nucl. Phys. {\bf B408} (1993) 512;\\
H.J. de Vega and A. Gonz\'{a}lez-Ruiz, Nucl. Phys. {\bf B417} (1994) 553.
%
\bibitem{devegahw}
H.J. de Vega and A. Gonz\'{a}lez-Ruiz, `Highest Weight Property for the
$SU_q(n)$ Invariant Spin Chains', LPTHE-PAR 94/13 (1994).
%
%
\bibitem{georg} G. J\"{u}ttner and M. Karowski, `The ``Good" Bethe
Ansatz Solutions of Quantum Group Invariant Heisenberg Models',
SFB 288 preprint (1994).
%
\bibitem{devegan} H.J. de Vega, Int. J. Mod. Ph. A {\bf 4} (1989) 2063.
%
\bibitem{resht} N. Yu. Reshetikhin `Quantized Univ. Envel. Algebras, YBE
and Invariants of Links (Part I)', LOMI E-4-87 (1987);\\
%
V. Pasquier and H. Saleur, Nucl. Phys. {\bf B330} (1990) 523;\\
%
N. Reshetikhin and V.G. Turaev, Invent. math. 103 (1991) 547;\\
J. Fuchs, `Affine Lie Algebras and Quantum Groups',
Cambridge Univ. Press (1992).
%
\bibitem{takahashi} M. Takahashi and M. Suzuki, Progr. Th. Ph., Vol. 48,
No. 6b (1972) 2187.
\end{thebibliography}
\end{document}